\documentclass[a4paper,11pt]{article}
\usepackage{pos}
\newcommand{\mw}{m_W}
\newcommand{\mwi}{m_{W,i}}
\newcommand{\pt}{p_t^\ell}
\newcommand{\mt}{M_t^{\ell\nu}}
\newcommand{\ptw}{p_t^{\ell\nu}}
\newcommand{\beq}{\begin{eqnarray}}
\newcommand{\eeq}{\end{eqnarray}}
\newcommand{\apt}{{\cal A}_{\pt}}
\renewcommand{\mid}{{\rm mid}}
\renewcommand{\d}{{\rm d}}

\title{A new observable for $W$-mass determination}

\author[a]{Luca Rottoli}
\author*[b]{Paolo Torrielli}
\author[c]{Alessandro Vicini}

\affiliation[a]{Physik Institut, Universit\"at Z\"urich, CH-8057 Z\"urich,
Switzerland}
\affiliation[b]{Dipartimento di Fisica, Universit\`a di Torino and INFN,
Sezione di Torino, I-10125 Torino, Italy}
\affiliation[c]{Dipartimento di Fisica, Universit\`a di Milano and INFN,
Sezione di Milano, I-20133 Milano, Italy}

\emailAdd{luca.rottoli@physik.uzh.ch}
\emailAdd{paolo.torrielli@to.infn.it}
\emailAdd{alessandro.vicini@mi.infn.it}

\abstract{In this contribution we discuss the properties of the jacobian asymmetry,
the new observable introduced in \cite{Rottoli:2023xdc} for a robust determination
of the value and uncertainty of the $W$-boson mass at hadron colliders.}

\FullConference{16th International Symposium on Radiative Corrections: Applications
of Quantum Field Theory to Phenomenology (RADCOR2023)\\
28th May - 2nd June, 2023\\
Crieff, Scotland, UK\\}

\begin{document}
\maketitle

\section{Introduction}
The determination of the $W$-boson mass $\mw$ is of paramount importance
for the precision programme of collider facilities such as the LHC
\cite{ATLAS:2017rzl,ATLAS:2023fsi,LHCb:2021bjt}.
In the Standard Model (SM), quantum corrections to the value of $\mw$ are
sensitive to other fundamental parameters of the theory, such as the top-quark
and the Higgs-boson masses. Therefore, $\mw$ is central to global fits of
electroweak (EW) precision observables \cite{Baak:2014ora,deBlas:2021wap},
allowing for compelling tests of the SM itself
\cite{Awramik:2003rn,Degrassi:2014sxa}.

Measurements of the $W$-boson mass at different high-energy colliders span
four decades, and the precision with which $\mw$ is determined has steadily
improved since discovery, owing to the wealth of available data and to many
experimental advances, reaching nowadays the level of 10-20 MeV
\cite{ATLAS:2017rzl,ATLAS:2023fsi,LHCb:2021bjt,CDF:2022hxs}, i.e.~a relative
accuracy at the permyriad level.
Such a level of precision requires an exquisite control over all elements that
feed into the extraction, including not only experimental calibrations, but
also the robustness of the strategies adopted to infer $\mw$ from data.
The consideration that the most precise $\mw$ measurement to date \cite{CDF:2022hxs}
is several standard deviations away from the SM expectations (and from the
World average) further stimulates a careful assessment of the methodologies
employed for $\mw$ determination.

At hadron colliders, the value of $\mw$ is primarily deduced from the
charged-current Drell-Yan (CCDY) process, the hadro-production of a
lepton-neutrino pair. Of particular relevance in this context are observables
defined in the transverse plane with respect to the collision axis, such as
the charged-lepton transverse momentum $\pt$, or the lepton-neutrino transverse
mass $\mt$ \cite{ATLAS:2017rzl,ATLAS:2023fsi,LHCb:2021bjt,CDF:2013dpa,CDF:2022hxs}.
The spectra of such quantities display a kinematical jacobian peak whose position
directly depends on the value of $\mw$, hence the shape of these distributions
in the peak region can be used as a privileged probe to extract the $W$-boson
mass.
A physical description of the shape of the jacobian peak requires to take into
account a variety of theoretical and experimental effects. On the theory side,
the peak arises at the boundary of the available charged-lepton phase space at
leading order in QCD, whence soft QCD radiation causes an integrable singularity
\cite{Catani:1997xc} in the spectrum around the peak in fixed-order perturbation
theory, which calls for all-order resummation of infrared (IR) enhanced QCD effects.
QED final-state radiation also significantly affects the shape of the peak, as
well as mixed QCD-EW effects in the case of $\pt$ \cite{CarloniCalame:2016ouw}.
On the experimental side, the peak is relatively stable under detector effects
for $\pt$, while it gets significantly smeared by the latter in the case of $\mt$,
owing to neutrino reconstruction, see e.g.~Figure 1 of \cite{D0:2013jba}.

\section{Standard $\mw$ determination}
A common procedure to extract of the $W$-boson mass at hadron colliders is
through template fitting. Theoretical template distributions for $\pt$ or $\mt$
are computed with different hypotheses $\mwi$ for the value of the $W$-boson
mass, and then compared to experimental data. A measure $\chi_i^2$ (be it a
true $\chi^2$ or a likelihood) is defined to quantify the distance between
the templates and the experimental spectra, and the $W$-mass value is determined
as the $\mwi$ corresponding to the minimum $\chi_i^2$. The main challenge with
this strategy is that the shape of the jacobian peak needs to be controlled with
a relative accuracy at the permille level, in order to resolve
$\Delta\mw/\mw\sim10^{-4}$ effects, while the current theoretical accuracy on
$\pt$ and $\mt$ spectra in CCDY is rather at the percent level
\cite{Chen:2022cgv,Chen:2022lpw,Neumann:2022lft}.

The procedure is restored by leveraging the availability of high-precision $p_t^Z$
data in neutral-current Drell Yan (NCDY). The tools used to produce the theoretical
template distributions, typically flexible but low-accuracy parton-shower event
generators, are calibrated to give the best description of such $p_t^Z$ data,
primarily by tuning the parameters of a non-perturbative (NP) model (e.g.~the
intrinsic $k_t$ of partons in the proton, or the shower cutoff scale $Q_0$).
The same tuning setup deduced in NCDY is then used to produce the templates of
$\pt$ and $\mt$ in CCDY, and, after tuning, the templates typically give rise to
reasonably low minimum $\chi_i^2$ values.

Although the template-fitting procedure is long-established, with the aim of
$10^{-4}$ relative accuracy on $\mw$, it is legitimate to raise methodological
concerns about its reliability and robustness. First, as evinced from the above
discussion, template fitting heavily relies on the tuning step of parton showers,
i.e.~the theoretical prediction is driven by NP physics, which is the least
understood theoretically.
In particular, tuned NP parameters effectively mimic the contributions of
higher-order perturbative radiation, and the significant progress in the
perturbative understanding of the Drell-Yan process
\cite{Duhr:2020sdp,Duhr:2021vwj,Chen:2021vtu,Chen:2022lwc,
Bizon:2018foh,Bizon:2019zgf,Ebert:2020dfc,Becher:2020ugp,Camarda:2021ict,
Ju:2021lah,Chen:2022cgv,Chen:2022lpw,Neumann:2022lft,
Dittmaier:2001ay,Baur:2001ze,Baur:2004ig,Arbuzov:2005dd,Zykunov:2005tc,
CarloniCalame:2006zq,CarloniCalame:2007cd,Arbuzov:2007db,Dittmaier:2009cr,
Balossini:2008cs,Balossini:2009sa,Barze:2012tt,Barze:2013fru,Dittmaier:2014qza,
Dittmaier:2015rxo,Bonciani:2016wya,deFlorian:2018wcj,Bonciani:2019nuy,
Delto:2019ewv,Cieri:2020ikq,Bonciani:2020tvf,Buccioni:2020cfi,Behring:2020cqi,
Buonocore:2021rxx,Bonciani:2021iis,Bonciani:2021zzf,Behring:2021adr,
Buccioni:2022kgy} is not fully exploited (barring to a certain extent effects
due to reweighing of event samples).
Moreover, the universality of the underlying NP model \cite{Konychev:2005iy},
assumed when applying to CCDY the same NP parameters extracted from NCDY,
can be spoiled by a variety of effects that differ in the two Drell-Yan
processes
\cite{Berge:2005rv,Pietrulewicz:2017gxc,Bagnaschi:2018dnh,Bacchetta:2018lna},
driven by the different parton flavour combinations they probe.
Finally, the definition of $\chi^2$ used as a distance measure does not
include theoretical uncertainty, owing to the non-statistical nature of
unphysical-scale variations. These features expose the template-fitting
procedure to a potential severe underestimation of the real uncertainties
associated with $\mw$ extraction.

From the above considerations, it would be desirable to define a procedure of
$\mw$ determination that allows for a transparent discussion of theoretical
and experimental uncertainties on the extracted $W$-boson mass value, and one
which would ideally minimise the reliance on $p_t^Z$ calibration/tuning when
extracting $\mw$.

\section{New strategy for $\mw$ determination}
\begin{figure}[t!]
\begin{center}
\includegraphics[width=0.45\textwidth]{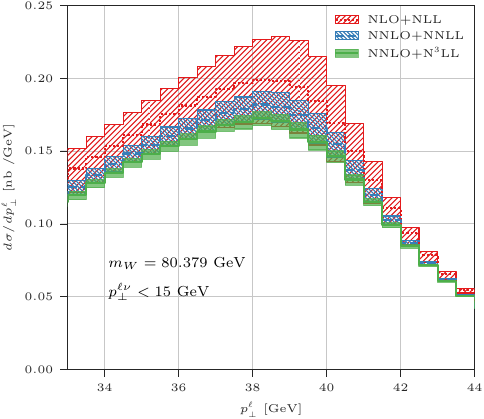}
\qquad
\includegraphics[width=0.45\textwidth]{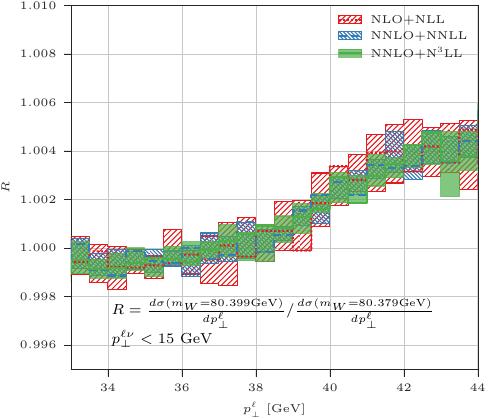}
\end{center}
\caption{\label{fig:ptlep}
Left: $\pt$ distribution in CCDY, computed with different QCD approximations
and reference $\mw=80.379$ GeV. Right: ratio of $\pt$ distributions computed
with two $\mw$ values differing by 20 MeV. Setup employed: $W^-$ production
at the 13-TeV LHC with acceptance cuts $\pt>20$ GeV,\, $\mt > 27$ GeV,\,
$|\eta_\ell| <2.5$,\, 66 GeV $< M^{\ell\nu} < 116$ GeV, $\ptw < 15$ GeV
($\eta_\ell$ and $M^{\ell\nu}$ being the charged-lepton rapidity and the
lepton-pair invariant mass, respectively); central replica of the NNPDF4.0
NNLO proton PDF set~\cite{NNPDF:2021njg} with strong coupling constant
$\alpha_s(m_Z) = 0.118$ through the LHAPDF interface~\cite{Buckley:2014ana};
$\ptw$ resummation and matching provided by RadISH \cite{Monni:2016ktx,Bizon:2017rah,Monni:2019yyr,Re:2021con},
fixed-order predictions provided by MCFM \cite{Campbell:2019dru}.
}
\end{figure}

To define the advocated novel procedure, we focus on the $\pt$ spectrum for
definiteness. This is displayed in Figure~\ref{fig:ptlep} (left) at various
perturbative accuracies, where one can appreciate the physical description
of the jacobian peak at $\pt\sim\mw/2$ provided by QCD resummation (i.e.~the
absence of any integrable singularity), as well as the $O(2\%)$ theoretical
accuracy achieved with state-of-the-art tools. The setup employed to obtain
the plot is detailed in the caption of the Figure. The right panel of
Figure~\ref{fig:ptlep} shows the remarkable feature that the ratio of $\pt$
spectra obtained with different $\mw$ values is largely independent of the
underlying QCD accuracy (provided resummation is included in the prediction),
which can be understood since the sensitivity to the value of $\mw$ stems from
$W$-boson propagation and decay, and it is essentially factorised from QCD
initial-state radiation. The sensitivity to a relative variation
$\Delta\mw/\mw\sim 2 \cdot 10^{-4}$ is also well resolvable beyond the
theoretical scale-uncertainty band (the latter being obtained as a 9-point
envelope varying renormalisation, factorisation and resummation scales by
factors of 2 around their central values).
%
 
\begin{figure}[t!]
\begin{center}
\includegraphics[width=1\textwidth]{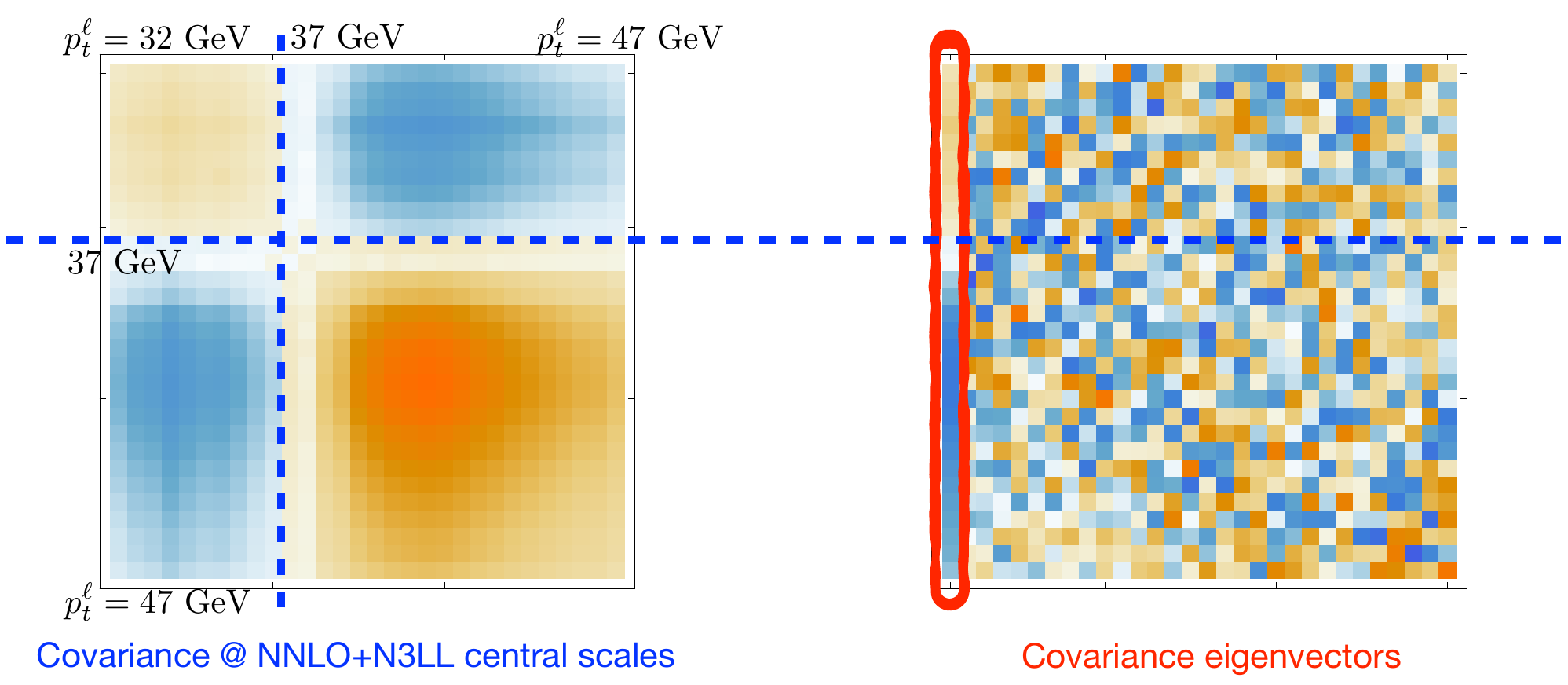}
\end{center}
\caption{\label{fig:cov}
Left: covariance matrix $C_{ij}^{(\mw)}$ at NNLO+N$^3$LL with central scales,
see main text for its detailed definition. Right: matrix of $C_{ij}^{(\mw)}$
eigenvectors as columns, sorted from left to right by decreasing eigenvalue
magnitude.}
\end{figure}

In order to quantify which of the $N$ bins $\sigma_i$ carry most of the
sensitivity to $\mw$, we can define a covariance matrix w.r.t.~$\mw$
variations as ${\cal C}_{ij}^{(\mw)} = \langle \sigma_i \, \sigma_j \rangle -
\langle \sigma_i \rangle \, \langle \sigma_j \rangle$, where $\langle x
\rangle = \frac1p \sum_{k=1}^p x_{(k)}$ denotes an average over the
range of $p$ different $\mw$ hypotheses.
By diagonalising ${\cal C}_{ij}^{(\mw)}$ one gets $N$ orthogonal $\pt$-bin
combinations whose eigenvalues represent the sensitivity of such eigenvectors
to $\mw$ variations. The left panel of Figure~\ref{fig:cov} displays for
illustrative purposes the covariance ${\cal C}_{ij}^{(\mw)}$ for NNLO+N$^3$LL
predictions with central scales, from which one can appreciate a clear
anti-correlation between the $\pt$ regions at the left and at the right of
the jacobian peak (orange bins are positive, blue bins are negative, their
magnitude being proportional to colour intensity). The right panel of
Figure~\ref{fig:cov} shows the matrix of eigenvectors of ${\cal C}_{ij}^{(\mw)}$,
sorted from left to right by decreasing eigenvalues.
%
Only the leftmost eigenvector displays a clear pattern of coefficients (positive
for $\pt$ below 37 GeV, negative above). Its corresponding eigenvalue $e_1$
is by far the dominant one, with $e_1/{\rm tr}[{\cal C}_{ij}^{(\mw)}]\sim 0.99$.
This means that most of the sensitivity to $\mw$ variations hiding in the $\pt$
spectrum is captured by the sole first bin combination.

On physical grounds, this pattern lends itself to a relatively straightforward
interpretation: a variation $\Delta\mw$ in the $W$-boson mass has the sole (or
at least by far dominant) effect of inducing a rigid shift of the $\pt$ spectrum
by $\Delta\mw/2$. A single $\pt$-bin combination is thus sufficient to entirely
capture the effect of $\mw$ variations. Such a combination is the one corresponding
to the translational mode of the spectrum: indeed its coefficients are proportional
to the derivative of the spectrum w.r.t.~$\pt$ (i.e.~$\d^2 \sigma/\d
{\pt}^2$), as expected from the generator of $\pt$ translations.

This analysis suggests two possible strategies. One would be that of measuring
and template-fitting directly $\d^2 \sigma/\d{\pt}^2$, as opposed to
$\d\sigma/\d\pt$: the former essentially distills all of the sensitivity to
$\mw$ variations, without being blurred by other concurring effects, chiefly
QCD radiation (cfr. independence of the spectrum derivative from QCD in the
right panel of Figure~\ref{fig:ptlep}). Whether this strategy is more robust
and resilient to tuning than the standard fitting procedure is subject of
future developments.
Another strategy, pursued in \cite{Rottoli:2023xdc}, is to encode (as much as
possible of) the information carried by the translational eigenvector into the
definition of a simple observable enjoying good features both from the
experimental and from the theoretical point of view.

\section{Jacobian asymmetry}
The dominant translational eigenvector of ${\cal C}_{ij}^{(\mw)}$ collects the bins
below (above) the peak with positive (negative) coefficients. An observable encoding
this information is thus the \emph{jacobian asymmetry} ${\cal A}_{\pt}$, defined as
\beq
&&
\apt({\pt}_{,\min}, {\pt}_{,\mid}, {\pt}_{,\max})
\, = \,
\frac{L-U}{L+U}
\, ,
\nonumber\\[5pt]
&&
L
\, = \,
\int_{{\pt}_{,\min}}^{{\pt}_{,\mid}} 
\d\pt \, \frac{\d\sigma}{\d\pt}
\, ,
\qquad
U
\, = \,
\int_{{\pt}_{,\mid}}^{{\pt}_{,\max}} 
\d\pt \, \frac{\d\sigma}{\d\pt}
\, ,
\eeq
where three values ${\pt}_{,\min}$, ${\pt}_{,\mid}$, and ${\pt}_{,\max}$
define two adjacent windows in the $\pt$ spectrum, selecting bins below
and above the peak, respectively. Crucial is that ${\pt}_{,\mid}$ be close
to 37 GeV, in order to match the change of sign in the coefficients of the
covariance eigenvector (see Figure~\ref{fig:cov}) \footnote{Alternatives
to $\apt$ could be devised: for instance, the two $\pt$ windows might not
be exactly adjacent, or one could give relative weights to $L$ and $U$,
in order to better align $\apt$ to the dominant covariance eigenvector.}.

A feature one can immediately notice in this definition is that $\apt$ is
constructed as a combination of fiducial rates in relatively wide $\pt$
windows (imagining ${\pt}_{\min} \sim 30$ GeV, and ${\pt}_{\max} \sim 50$
GeV), whence $\apt$ is a single scalar number experimentally measurable
by means of an inclusive counting of events in the two windows.

\begin{figure}[t!]
\begin{center}
\includegraphics[width=0.5\textwidth]{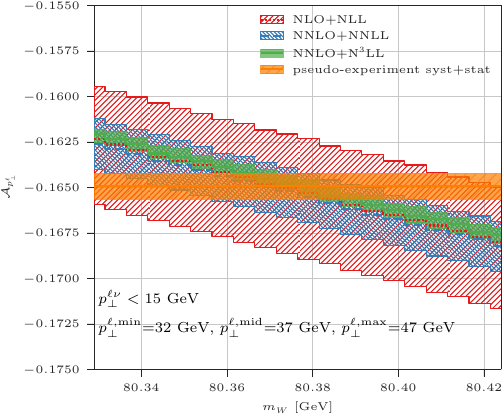}
\end{center}
\caption{\label{fig:asy}
Jacobian asymmetry as a function of $\mw$ at different perturbative QCD orders.
Window edges are specified in the plot. See text for details.}
\end{figure}

Figure~\ref{fig:asy} displays $\apt$ in various QCD approximations, as a
function of $\mw$. Its linearly decreasing behaviour stems from the fact
that an $\mw$ shift by $+\Delta\mw$ induces a shift in the position of the
jacobian peak by $+\Delta\mw/2$ (i.e.~linear in $\Delta\mw$), depleting
$L$ and populating $U$ if ${\pt}_{\mid}$ is at the left of the peak.
The slope of $\apt$ as a function of $\mw$ is independent of the QCD
approximation and of the scale choice, which again reflects the factorisation
of QCD initial-state radiation from the $\mw$-sensitive propagation and
decay; this feature carries over to NP QCD effects as well \cite{Rottoli:2023xdc},
which will just result in a rigid shift of $\apt$. The slope itself is
related to the magnitude of the first covariance eigenvalue, and depends
on the value of the chosen window edges.
%

From Figure~\ref{fig:asy} one can appreciate the excellent perturbative-QCD
convergence properties of the observable, where predictions for $\apt$ at
higher orders perfectly lie within lower-order uncertainty bands, with a
residual theoretical uncertainty steadily decreasing while including more
accurate predictions. This ultimately highlights the importance of
state-of-the-art results for high-accuracy Drell Yan predictions.
The perturbative merits of $\apt$ are not unexpected, since the observable
is inclusive over radiation in wide $\pt$ windows. This very feature has
also evident experimental advantages. On one side, the measurement of $L$
and $U$ should be relatively simple, with both systematic and statistical
uncertainties under good control; on the other side, the usage of wide
fiducial windows should be beneficial towards unfolding detector effects,
allowing for a combination of different $\mw$ determinations
\cite{Amoroso:2022rly,Amoroso:2023pey}.
The orange band in Figure~\ref{fig:asy} reports the results of a putative
experimental measurement of $\apt$: the central value is arbitrary, while
the uncertainty band is obtained by propagating in quadrature a realistic
$0.1\%$ systematic error on $L$ and $U$, and assuming no correlations (the
statistical error is already negligible with a luminosity ${\cal L} = 140$
fb$^{-1}$).

In the context of $\apt$, the $W$-boson mass would just be extracted as the
intersection of two non-parallel straight bands, making it straightforward
both to include new theoretical and experimental refinements, as they become
available, and to interpret robustly the effect and the uncertainty of each
of the various contributions to the observable, e.g.~the impact of different
PDF choices, of NP QCD contributions, of EW corrections, of detector effects,
and so on.
As a consequence, the procedure of extracting $\mw$ through $\apt$ would not
be entirely driven by tuning to NCDY data, as the NP contribution would just
become one of the many ingredients (and most probably of modest impact)
concurring to an accurate prediction of $\apt$, thus of the $W$-boson mass.
%

\begin{figure}[t!]
\begin{center}
\includegraphics[width=0.45\textwidth]{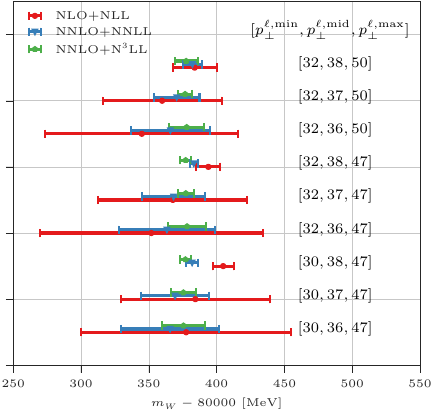}
\qquad
\includegraphics[width=0.45\textwidth]{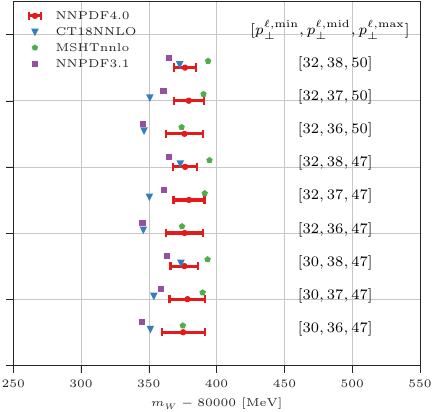}
\end{center}
\caption{\label{fig:asy2}
Left: jacobian asymmetry with various choices for window edges. Right: study of the
impact of PDFs on the jacobian asymmetry.
}
\end{figure}
In the left panel of Figure~\ref{fig:asy2} we display results for $\apt$ for different
choices of the window edges ${\pt}_{\min}$, ${\pt}_{\mid}$, and ${\pt}_{\max}$.
In general the perturbative convergence is very well behaved across different values
of the edges, and we stress the importance of N$^3$LL resummation also for assessing
the quality of the perturbative convergence, and to check it beyond the mere level
of scale variations. We notice a certain trade-off between sensitivity to $\mw$
variations, improving at higher ${\pt}_{\mid}$, and perturbative stability, improving
at lower ${\pt}_{\mid}$.
Given the general convergence pattern, we conclude that an $\mw$ determination with
a perturbative-QCD accuracy $\Delta\mw\sim\pm 5$ MeV seems achievable by means of
$\apt$.

The right panel of Figure~\ref{fig:asy2} shows the impact on $\apt$ due to considering
the envelope of the PDF replicas within our default set, or to employing alternative
PDF sets \cite{Hou:2019efy,Bailey:2020ooq,NNPDF:2017mvq}.
In the former case, a spread of $\Delta\mw\sim\pm12$ MeV is induced using the NLO+NLL
result as a baseline; varying central PDF set conversely results in an $O(30$ MeV)
effect on $\mw$ at NNLO+N$^3$LL. Such effects can however be reduced to few MeV via
PDF profiling, employing further information not included in $\apt$, such as additional
$\pt$ bins \cite{Bagnaschi:2019mzi}, the anti-correlation of different rapidity windows
\cite{Bozzi:2015hha,Bozzi:2015zja}, the combination of $W^+$ and $W^-$ production
channels \cite{ATLAS:2017rzl}.

As a concluding remark, if the exercise leading to the left panel of
Figure~\ref{fig:asy2} is repeated in the operating conditions of the CDF II experiment,
the theoretical-QCD uncertainty associated to $\mw$ by means of the jacobian asymmetry
is $\sim\pm 30$ MeV ($\sim\pm 10$ MeV) using $\pt$ ($\mt$) at NLO+NNLL, which is the
accuracy of the theoretical tools
\cite{Balazs:1997xd,Landry:2002ix,Isaacson:2022rts}
employed by CDF II. Such figures should be compared to the $O(\pm2$ MeV)
perturbative-model uncertainties quoted by the CDF measurement \cite{CDF:2022hxs}.

\section{Outlook}
The study of theoretical uncertainties in the context of $W$-mass extraction is
crucial, aiming at $10^{-4}$ relative accuracy, as well as particularly
delicate.
The standard procedure of template fitting makes this study involved, especially
owing to a necessary step of calibration of the predictions to neutral-current
Drell-Yan data, which are to a certain extent extraneous to the $W$-boson production
process: tuning to data certainly improves the accuracy of the description, but
not the precision of the underlying physics model.

The proposed jacobian asymmetry $\apt$, described in this contribution, represents an
attempt to cast the discussion of $\mw$ uncertainties on more solid theoretical grounds.
While capturing most of the sensitivity to $\mw$ variations, thereby representing a
good candidate observable for $\mw$ determination, $\apt$ displays remarkable
perturbative-QCD properties of accuracy and stability. It also allows good control
over statistical and systematic experimental errors, and it allows for a simple
unfolding of detector effects, in view of a global combination of different experimental
$\mw$ measurements.

A solid prediction for $\mw$ through $\apt$ necessarily hinges upon a reliable
description of all effects concurring to the latter, such as PDF variations,
non-perturbative modelling, EW and mixed QCD-EW radiation, higher orders in QCD,
together with a concrete assessment of the associated uncertainties.
While the predictions for $\apt$ presented above are still partial, in that they
lack most of these input theoretical ingredients, a new roadmap for $\mw$ measurement
is set up where each of these effects can be separately included and assessed in
detail, helping reaching a general theory-experiment consensus on the value and
the uncertainties associated to the $W$-boson mass.

\section*{Acknowledgements}
LR is supported by the Swiss National Science Foundation contract PZ00P2\_201878.
PT has been partially supported by the Italian Ministry of University and Research
(MUR) through grant PRIN 20172LNEEZ and by Compagnia di San Paolo through grant
TORP\_S1921\_EX-POST\_21\_01.
AV is supported by the Italian MUR through grant PRIN 201719AVICI\_01.

\bibliographystyle{abbrv}

\end{document}